\documentclass[pra,aps,twocolumn,amsmath,amssymb,showpacs,preprintnumbers]{revtex4}
\usepackage{graphicx}
\usepackage{dcolumn}
\usepackage{bm}
\usepackage{latexsym}
\usepackage{amssymb}
\usepackage{amsmath}
\usepackage[mathscr]{eucal}
\usepackage{enumerate}
\newcommand{\matA}[4]{
  \left[
    \begin{array}{cc}   
        #1 & #2 \\      
        #3 & #4
    \end{array}
  \right]}

\newcommand{\matAH}[2]{
  \left[
    \begin{array}{cc}   
        #1 & #2          
    \end{array}
  \right]}

\newcommand{\matAAH}[2]{
    \begin{array}{cc}    
        #1 & #2          
    \end{array}
}

\newcommand{\matAV}[2]{
  \left[
    \begin{array}{c}    
        #1 \\           
        #2
    \end{array}
  \right]}

\newcommand{\matAAV}[2]{
    \begin{array}{c}     
        #1 \\            
        #2
    \end{array}
}

\newcommand{\tf}[4]{
\left[
  \begin{array}{c|c}
      #1 & #2 \\ \hline
      #3 & #4
  \end{array}
\right]}

\newcommand{\matAA}[4]{
  \begin{array}{cc}
      #1 & #2 \\ 
      #3 & #4
  \end{array}
}

\newcommand{\matB}[9]{
  \left[
    \begin{array}{ccc}  
        #1 & #2 & #3 \\ 
        #4 & #5 & #6 \\ 
        #7 & #8 & #9
    \end{array}
  \right]}

\newcommand{\matBV}[3]{
  \left[
    \begin{array}{c}    
        #1 \\           
        #2 \\           
        #3 
    \end{array}
  \right]}

\newcommand{\matBH}[3]{
  \left[
    \begin{array}{ccc}  
        #1 & #2 & #3 
    \end{array}
  \right]}

\newcommand{\matBBH}[3]{
    \begin{array}{ccc}  
        #1 & #2 & #3 
    \end{array}
  }

\newcommand{\wig}[6]{
  \left\{
    \begin{array}{ccc}  
        #1 & #2 & #3\\  
        #4 & #5 & #6
    \end{array}
  \right\}}

\newcommand{\mean}[1]{\langle{#1}\rangle}

\newcommand{\bra}[1]{\langle{#1}|}
\newcommand{\ket}[1]{|{#1}\rangle}

\newcommand{\dgg}{^{\dag}}
\newcommand{\Tr}{{\rm Tr} \ }

\newcommand{\im}{{\rm i}}
\newcommand{\E}{{\rm E}}

\newcommand{\maths}[1]{\mathscr{#1}}
\newcommand{\p}{\partial}
\newcommand{\cg}[3]{{\rm C}_{#1 \ #2}^{#3}}

\begin{document}

\title{Control Theoretical Approach to Quantum Control}

\author{Masahiro Yanagisawa}
\email{yanagi@cds.caltech.edu}
    \affiliation{
Control and Dynamical Systems, California Institute of Technology,
Pasadena CA 91125
}

\date{\today}

\begin{abstract}
 We derive the quantum stochastic master equation for bosonic systems
 without measurement theory but control theory.
 It is shown that the quantum effect of the measurement can be represented
 as the correlation between dynamical and measurement noise.
 The transfer function representation allows us to 
 analyze a dynamical uncertainty relation which imposes strong
 constraints on the dynamics of the linear quantum systems.
 In particular, quantum systems preserving the minimum uncertainty are
 uniquely determined.
 For large spin systems, it is shown that local dynamics are equivalent
 to bosonic systems. 
 Considering global behavior, we find quantum effects to which there is
 no classical counterparts.
 A control problem of producing maximal entanglement is discussed as the
 stabilization of a filtering process.
\end{abstract}

\pacs{03.65.Ta,42.50.Lc,02.30.Yy}

\maketitle

\section{Introduction}
\label{in}

The quantum stochastic master equation has been derived by analyzing a
measurement process carefully \cite{car,wis1} 
in combination with the notion of positive operator valued measure
\cite{kra}.
Since it is based on the model that the measurement is made on the quantum
system indirectly via an environment, the stochastic master equation is
deeply involved in the input-output formulation \cite{gar,col}, 
which was introduced as a response to the physical necessity for the
formulation of quantum system interacting the traveling field.
This formulation is compatible with control theory and extensively
analyzed using the transfer function representation \cite{yanagi1}.
Due to the control theoretical analysis, the full quantum treatment of
feedback system and controller design was developed
for quantum noise reduction problems \cite{yanagi2}.

The quantum stochastic master equation for bosonic fields has a decent
property that the gaussianity of a density matrix is preserved in time.
Thus, the full dynamics of the density matrix can be described by up to
the second moments \cite{wis2,yanagi3,jac}, 
and the covariance matrix of the canonical observables obeys the matrix
analog of second order equation, the Riccati differential equation, 
which is widely known in control theory such as 
linear quadratic Gaussian control, estimation, 
$H_2$ and $H_\infty$ control \cite{doy,kim}. 
These formulations of the quantum stochastic master equation is, 
however, based on the Hamiltonian formulation and incorporated with
physical treatments, and therefore sometimes misled engineers about 
the back-action of the quantum measurement. 

In this paper, we derive the stochastic master equation from a general
description of linear quantum systems via the transfer function
representation \cite{yanagi1}.
In this treatment, we do not assume any underlying physics and
Hamiltonians.
The only assumption imposed here is the noncommutativity of the input
and the output signals. 
Although all parameters of the linear quantum system are left
undetermined, the noncommutativity leads to strong constraints on the
transfer function and allows us to obtain the equivalent model to that
of physics.
These constraints are characterized by the fact that 
the poles and zeros of the transfer function are distributed 
in the complex plane subject to a certain rule. 
This includes significant information when one designs a quantum system
because the zero and pole of the transfer function completely determine
the statistical properties of the output.
Then, the stochastic master equation can be derived as a conditional
density of the linear system with the constraints. 
The quantum mechanical constraints can also be characterized by 
the detectability condition via the Hamiltonian matrix associated with
the algebraic Riccati equation.

These ideas can be extended to spin systems using the quasiprobability
distribution function on the sphere \cite{sto}.
At first we will show the fermionic analog of the relation between
the superoperators and the differential operators by means of the
so-called star product.
This relation allows us to establish the same formalism as the bosonic 
case for the local behavior of a large spin.
The global behavior of the large spin is considered for quantum
non-demolition (QND) measurement.
This formulation of the spin system shows that the leading expansion
with respect to the spin number has the corresponding classical system.
Dealing with higher order expansion, we will see 
a quantum effect of measurement in the back-action process 
to which there are no classical counterparts.

For experimental and practical reasons, spin systems are thought of as 
an important basis for quantum information processing, and controlling
spin systems is a indispensable technology for it.
In particular, spin entanglement is an important subject of quantum
control and a lot of experiments have been performed using QND
measurement in recent years \cite{kuz,sto1,sto2}.
This problem can be reduced to the stabilization of the spin system with
a stochastic noise.
We will consider the production of the maximal entanglement with the use
of feedback based on the global description of the large spin system.

This paper starts with Sec.\ref{lt} which is a brief review of classical
linear system theory for introducing control theoretical notions to
quantum systems. 
In Sec.\ref{kf} and \ref{hm}, filtering theory is overviewed.
Sec.\ref{qm} introduce the general formulation of linear quantum systems,
and the uncertainty relation is stated in terms of the transfer function
in Sec.\ref{utf}.
We formulate measurement on linear quantum systems in Sec.\ref{mp}
through which the uncertainty relation is stated again in terms of
detectability in Sec.\ref{sur}.
The mathematical basis for the spin system is developed in Sec.\ref{bswf}.
The local behavior and the global behavior of the spin system are
considered in Sec.\ref{lbls} and \ref{ghos}, respectively.
The production of the spin entanglement based on the
global behavior is discussed in Sec.\ref{dis}.

\section{linear system theory}
\label{lt}

A linear system is described in the time domain as
\begin{subequations}
\label{le}
\begin{align}
 \dot{x}(t) &= Ax(t) + Bu(t), \\
 m(t) &= Cx(t) + Du(t),
\end{align}
\end{subequations}
where vectors $u,m$ and $x$ represent the input, the output, and the state 
of the system, respectively.
The linearity of the system allows us to express it in the frequency
domain as 
\begin{align}
  m(s)=G(s)u(s),
\label{ito}
\end{align}
where $u(s),m(s)$ are the Laplace transforms of the input and the output.
A function $G(s)$ relating $u(s)$ to $m(s)$ is called
a {\it transfer function}, defined as 
\begin{align}
  G(s) = C(s-A)^{-1}B+D := \tf{A}{B}{C}{D}.
\end{align}
Each element of the transfer function is referred to as $A,B,C$ and $D$ 
matrices, respectively.

The transfer function representation of the linear system 
shows that the state $x$ is a hidden information carrier from the
input to the output,
and the choice of $x$ is not important in the relation between
them. 
The input-output relation is invariant under the similarity
transformation of $x$, i.e., for any nonsingular matrix $T$, 
\begin{align}
 \tf{A}{B}{C}{D} = \tf{TAT^{-1}}{TB}{CT^{-1}}{D}.
\label{stf}
\end{align}
For the same reason, the stability is characterized by 
an invariant quantity under the similarity transformation.
The system is said to be {\it stable} if $\mbox{Re}\lambda(A)<0$.
$\lambda(A)$ denotes the eigenvalues of $A$, 
which are called the {\it poles} of $G$.

$B$ matrix determines the relationship between the state of the system and 
the input, and thereby involves in controllability of the system.
The pair $(A,B)$ is said to be {\it controllable} if, for any set of complex
numbers $\Lambda=\{\lambda_1,\cdots,\lambda_n\}$, there exist a matrix
$K$ such that $\lambda(A+BK)=\Lambda$.
This definition indicates that the poles of $G$ are assignable anywhere
in the complex plane ${\bf C}$ by means of state feedback $u=Kx$. 
The pair $(A,B)$ is said to be {\it stabilizable} 
if there exist a matrix $F$ such that $A+BF$ is stable.
This is the case if uncontrollable modes are stable.
Likewise, $C$ matrix determines which element of the state is 
visible for an observer.
The pair $(C,A)$ is said to be {\it observable} if $(A^T,C^T)$ is
controllable, and {\it detectable} if there exists $L$ such that $A+LC$ is
stable.

In case of single-input-single-output (SISO) systems, 
due to $D$ matrix,
there exists an $s$ on the complex plane for which 
$G=0$.
Such a point is called the {\it zero} of $G$ and 
plays an important role in a wide class of control problems
as a ``norm'' of the system is determined by both of 
the pole and zero of $G$.
For example, 
if the input is a noise and the output is a quantity to be not 
affected by the noise, then 
one would design the system to have a small norm by appropriately 
distributing the poles and zeros on ${\bf C}$.

\section{filtering}
\label{kf}

The notion of observability is concerned with the determinacy
of the initial state $x(0)$ from the output segment 
$\{m(s)|0\le s < t\}$ of arbitrary length for a deterministic
measurement. 
For a noisy measurement, a filtering process is required to extract
information about the system subject to a certain criterion.

\subsection{uncorrelated noises}

Let us consider a nonlinear system 
with two independent noises represented as 
\begin{subequations}
\label{ls}
\begin{align}
 dx &= a(x)dt + b(x)dw \\
 dm &= c(x)dt + Ddv
\end{align}
\end{subequations}
where $w,v$ are independent normalized Wiener processes, and we have
assumed that $D$ is a constant matrix.
The first equation describes the dynamics of $x$ driven by the
stochastic signal $v$, and the second one is the noisy measurement
process. 
This system has the joint density $p(x,m)$ of two random variables $x,m$, 
and only $m$ is directly visible. 
A filtering problem is to find an optimal estimate of $x$, $\hat{x}$, 
from the measurement outcome $m$.
If we define the optimality by minimization of the mean square error
$P=\E[(x-\hat{x})(x-\hat{x})^T]$,
then the estimate is given by a {\it conditional expectation}.
For an arbitrary function $\phi(x)$, 
the conditional expectation 
\begin{align}
\pi_t(\phi) := \E[\phi|m(s) \ (0\le s \le t)]
\nonumber
\end{align}
satisfies  
\begin{align}
  d\pi_t(\phi) =& \pi_t(\maths{L}\phi)dt 
  +[\pi_t(\phi c^T)
-\pi_t(\phi)\pi_t(c^T)](DD^T)^{-1}
d\underbar{v}
\label{nlf} 
\end{align}
where 
\begin{align}
 \maths{L} = \sum_i a_i\frac{\p}{\p x_i}
 + \frac{1}{2}\sum_{ijl}b_{il}b_{jl}\frac{\p^2}{\p x_ix_j},
\nonumber
\end{align}
and $d\underbar{v} := dm-\pi_t(c)dt$, which is called 
the {\it innovation process}, is another Wiener process.
Note that if $DD^T$ is singular, the inverse can be
replaced by the pseudoinverse.

From the conditional expectation (\ref{nlf}), 
it can be easily seen that the conditional density 
\begin{align}
 p(x,t) = \mbox{Pr}[x(t)|m(s) \ (0\le s \le t)]
\nonumber
\end{align}
satisfies 
\begin{align}
 dp = \maths{L}^* pdt
 + (c-\hat{c})^T(DD^T)^{-1}p \ d\underbar{v},
\label{cd} 
\end{align}
where $\maths{L}^*$ is the adjoint operator of $\maths{L}$, 
or the Fokker-Planck operator.

In the linear case,
$a(x)=Ax, \ b(x) = B, \ c(x)=Cx$,
the mean $\hat{x}$ and covariance matrix $P$ 
with respect to the conditional density $p$ are given by
\begin{align}
 d\hat{x} &= A\hat{x}dt + PC^T(DD^T)^{-1} d\underbar{v},
\label{nc} \\
 \dot{P} &= AP + PA^T +BB^T - PC^T(DD^T)^{-1}CP.
\nonumber
\end{align}
where $d\underbar{v}=dm-C\hat{x}dt$ is the innovation process.
This is known as the Kalman filter.
The second equation is referred to as the Riccati differential equation.

\subsection{correlated noises}

Let us consider a linear system with 
two normalized Wiener processes that are correlated as
\begin{align}
 \mean{dv(s)dw^T(t)}=Sdt\delta(t-s).
\end{align}
In this case, 
the conditional expectation is given by the Kalman filter 
\begin{align}
 d\hat{x} &= A\hat{x}dt + (PC^T+BS)(DD^T)^{-1} d\underbar{v},
\label{cr} \\ 
 \dot{P} &= AP + PA^T +BB^T 
\\  & \hspace*{1em}
  -(PC^T+BS)(DD^T)^{-1}(PC^T+BS)^T.
\nonumber
\end{align}
As in the uncorrelated case,
the conditional density is derived as the dual expression of the 
Kalman filter, given by
\begin{align}
 dp = \maths{L}^* pdt 
 +\Bigl[C(x - \hat{x}) -\nabla BS \Bigr]
 (DD^T)^{-1} p \ d\underbar{v}.
\label{cc}
\end{align}
Note that the correlation of the two noises appears as 
a derivative of first order in the coefficient of the innovation
process.

\section{Hamiltonian matrix}
\label{hm}

It can be shown that under the measurement for an infinitely long time,
the Kalman filter is in a stationary state if the stability is assured.
(The convergence also holds even if the system is not stable, provided
certain conditions are imposed.) 
Then, the Riccati differential equation for the covariance matrix is
reduced to the algebraic Riccati equation, which is deeply related to
the Hamiltonian matrix.

A matrix $H$ is said to be {\it Hamiltonian} if 
\begin{align}
 \Sigma H + H^T\Sigma = 0,
\label{dha}
\end{align}
where 
\begin{align}
 \Sigma = \matA{0}{-I}{I}{0}.
\nonumber
\end{align}
From this definition, it can be easily seen that the
Hamiltonian matrix is of the form
\begin{align}
 H=\matA{A}{R}{Q}{-A^T}.
\end{align}
where $W,Q$ are symmetric matrices.
Corresponding to this matrix, let us define an algebraic Riccati
equation as
\begin{align}
 XA+XA^T+XRX-Q=0.
\label{ric}
\end{align}
Let us denote by $\mbox{Ric}[H]$ a solution $X$ satisfying
$\mbox{Re}\lambda(A+RX)< 0$.

From Eq.(\ref{ric}), it follows that
\begin{align}
 \matA{I}{0}{-X}{I}H\matA{I}{0}{X}{I} = \matA{A+RX}{R}{0}{-(A+RX)^T},
\nonumber
\end{align}
which implies that 
$\lambda(H)=\lambda(A+RX)\cup \lambda(-A-RX)$.
Using Eq.(\ref{ric}) again, we have
\begin{align}
 H\matAV{I}{X} = \matAV{I}{X} (A+RX).
\label{arim}
\end{align}
It turns out that $\mbox{Ric}[H]$ is given by $X$ satisfying
Eq.(\ref{arim}) with $\mbox{Re}\lambda(A+RX)< 0$.
For $H$ which has eigenvalues on the imaginary axis, it is basically
possible to obtain a solution of the Riccati equation due to the same
procedure. 
This is the case if a quantum system is under QND measurement, as will
be seen later.

For an algebraic Riccati equation of the form
\begin{align}
 AX+ XA^T +BB^T - XC^TCP = 0,
\nonumber
\end{align}
the corresponding Hamiltonian matrix is written as
\begin{align}
 H = \matA{A^T}{-C^TC}{-BB^T}{-A}.
\nonumber
\end{align}
In this case, there exists $\mbox{Ric}[H]$ iff $(C,A)$ is detectable and
$(A,B)$ has no uncontrollable modes on the imaginary axis. 
Moreover, if these conditions hold, $\mbox{Ric}[H]\ge 0$.

\section{linear quantum system}
\label{qm}

In this section, we will introduce a basic formulation of linear quantum
systems. 
Let $u$ be a noncommutative variable such that 
\begin{equation}
  [u(t),u\dgg(t')]=\delta(t-t').
 \label{cu}
\end{equation}
The signal can be decomposed into the real and imaginary parts 
as
\begin{align}
 u_x =  u + u\dgg,
 \ \ \
 u_y = -\im(u-u\dgg).
\nonumber
\end{align}
Physically, these operators correspond to the amplitude of two phases of
a signal constituted of the infinite number of independent bosonic modes
in free space.

Let us consider a quantum system which converts the signal $u$ into 
another quantum signal $m(t)$ in the same space
according to a completely positive map $\Gamma$ given by
\begin{align}
 m(t) = \Gamma(u(t)) = u(t)*g(t),
\end{align}
where $g$ is a function determined by $\Gamma$, and $*$
represents the convolution of the operator $u$ and the function $g$.
$u,m$ can be thought of as the input and the output of the quantum
system. 
In the linear case,
an adequate rotation in the output complex amplitude plane 
can choose the form of the output operator $m$ 
such that $m_x,m_y$ respond to $u_x,u_y$, respectively
\cite{cav}.
Then, the two phases can be decoupled as
\begin{align}
\matAV{m_x}{m_p} = G \matAV{u_x}{u_y} 
                 =\matA{G_x}{0}{0}{G_y}\matAV{u_x}{u_y}, 
\label{df}
\end{align}
where $G_x,G_y$ are the transfer functions for each phase given by
\begin{align}
 G_x=
 \tf{A_x}
    {B_x}
    {C_x}
    {D_x},  \ \ 
 G_y=
 \tf{A_y}
    {B_y}
    {C_y}
    {D_y}.
\end{align}
In the time domain, this system can be represented as
\begin{subequations}
\label{dft}
\begin{eqnarray}
\hspace*{-4em}
 \matAV{dx}{dy} \hspace{-2mm} &=& \hspace{-2mm}
 \matA{A_x}{0}{0}{A_y}\hspace{-1mm}\matAV{x}{y}dt 
+ \matA{B_x}{0}{0}{B_y}\hspace{-1mm} \matAV{du_x}{du_p},
 \\ \hspace*{-3em}
 \matAV{dm_x}{dm_y} \hspace{-2mm} &=& \hspace{-2mm} 
 \matA{C_x}{0}{0}{C_y}\hspace{-1mm}\matAV{x}{y}dt 
+ \matA{D_x}{0}{0}{D_y}\hspace{-1mm} \matAV{du_x}{du_y},
\end{eqnarray}
\end{subequations}
where $x,y$ are the internal observables of the system.
They are noncommutative in general, however the explicit relation
between them is not necessary.
Thus, although we denote them by $x,y$, it does not necessarily means
that $x,y$ correspond to position and momentum.

The advantage of using transfer functions is that we can simplify the
expression of complex networks.
For example, the cascade connection of two quantum systems 
represented by $G_1,G_2$, 
which constitutes the simplest quantum network,
is described by the product of the two transfer functions 
\begin{align}
 G_1G_2 = \tf{\matAA{A_1}{B_1C_2}{0}{A_2}}
             {\matAAV{B_1D_2}{B_2}}
             {\matAAH{C_1}{D_1C_2}}
             {D_1D_2}.
\label{cas}
\end{align}
For any type of connections of systems, we can obtain a simple
expression using transfer functions.

\section{correlation and noise reduction}
\label{cnr}

Let us consider a quantum system in a density matrix 
$\rho$ on a Hilbert space $\maths{H}$.
To evaluate the correlation of noncommutative observables of the system
with respect to $\rho$, we introduce a pre-inner product of operators
$Q,R$ on $\maths{H}$ as \cite{hol}
\begin{align}
 \mean{Q,R}_{\rho} = \frac{1}{2}\Tr [\rho(QR\dgg+R\dgg Q)].
\end{align}

Let $q(t)$ and $r(t)$ be self-adjoint operators on ${\mathscr H}$,
and $q(\omega)$ and $r(\omega)$ be the corresponding Fourier transforms,
respectively.
The power spectrum $\maths{S}_{qr}$ of $q$ and $r$ is defined as
\begin{equation}
  \label{power}
  \mean{q(\omega),r(\omega')}_{\rho} 
  = \maths{S}_{qr}(\omega)\delta(\omega-\omega'),
\end{equation}
The correlation function $\maths{R}_{qr}(t)$ of operators $q(t)$ 
and $r(t)$ is then defined by inverse Fourier transform
of the power spectrum $\maths{S}_{qr}(\omega)$.

Returning to the decoupled system (\ref{df}),
it turns out that the absolute value of the transfer function
determines the power spectrum of the output signals, i.e.,
the power spectrum of $m_x$ is related to that of $u_x$ via
\begin{equation}
  \label{sgs}
  \maths{S}_{m_xm_x}(\omega) = 
  |G_x(\im \omega)|^2 \maths{S}_{u_xu_x}(\omega).
\end{equation}
If the transfer function $G_x$ is unitary,
then the power spectrums of
the input and output signals are equivalent.
Such a system cannot change the statistical property of the input
signal at all.
However, 
if we design the system such that the absolute value of the transfer
function $G_x$ is less than unity on the imaginary axis of the complex
plane ${\bf C}$, 
the fluctuation of the input is to be reduced through the system in
$x$-phase.
Requiring stability of $G_x$,
reducing the fluctuation of the input in x-phase
can be stated as 
\begin{align}
  |G_x(s)| < 1
  \label{sq}
\end{align}
in the left half plane of ${\bf C}$.
The output of the system satisfying the condition (\ref{sq}) 
is a squeezed state.
If $G_x$ has a zero on the origin in ${\bf C}$ then the system
can produce the perfect squeezing asymptotically.
This is the case if we use the quantum mechanical feedback and
parametric amplifier, no matter how weak the performance of the
amplifier is \cite{yanagi2}.

At this point,
it seems that each element of the transfer function can take an
arbitrary value, and there seems to be no differences between classical
and quantum systems. 
However, there is a very strong constraint on the transfer
function in the quantum case due to the noncommutativity of the input
and output signals, as will be seen in the next section.

\section{uncertainty and transfer function}
\label{utf}

The specific feature of quantum signals is the existence of
an additional skew-symmetric form. 
For arbitrary operators $Q,R$, we define \cite{hol}
\begin{align}
  [Q,R]_{\rho} = \im \Tr [\rho (R\dgg Q-Q R\dgg)].
  \label{skew}
\end{align}
Consider the SISO linear quantum system (\ref{df}) with an input signal
$u$ satisfying Eq.(\ref{cu}).
In the frequency domain, 
the input $u$ is characterized by 
\begin{align}
  [u_x(\omega),u_y(\omega')]_{\rho}
  =\frac{1}{4\pi}\delta(\omega+\omega').
  \label{comin}
\end{align}
It has been shown \cite{yanagi2} 
that the noncomutativity of the input and the output
result in a following relation between the two phases:
\begin{align}
  G_x(s)G_y(-s)\ge 1.
  \label{guncertain1}
\end{align}
If the output $m$ also satisfies the relation 
\begin{align}
  \label{comout}
  [m_x(\omega),m_y(\omega')]_{\rho}
  =\frac{1}{4\pi}\delta(\omega+\omega'),
\end{align}
then the equality of Eq.(\ref{guncertain1}) is achieved, i.e.,
\begin{align}
  G_x(s)G_y(-s)=1.
  \label{guncertain}
\end{align}
This is the case when the input and the output are in the same space.

The relation (\ref{guncertain})
allows us to state the quantum theoretical limitations
on linear quantum systems in terms of control theory.
Suppose that we design a quantum system such that 
$G_x$ is stable and 
has a zero on the origin in ${\bf C}$ to reduce the noise, as
stated in the previous section.
This system can produce the perfect squeezing asymptotically.
Then, according to the relation (\ref{guncertain}), the transfer
function of the other phase, $G_y$, has a pole at the origin, 
and consequently, $G_y$ is unavoidably unstable and the noise is amplified
to infinity. 
That is to say, in the quantum case, the zeros of $G_x$
and the poles of $G_y$ are distributed symmetrically with respect to the
origin in ${\bf C}$.
And also, the trade-off between the two phases can be easily seen by
taking the norm of the both side of Eq.(\ref{guncertain1}), i.e., 
$|G_x||G_y|\ge 1$.

It is worth noting that from Eqs.(\ref{sgs},\ref{guncertain}), 
the stochastic properties of the output is unconcerned with 
the noncommutativity of the internal observables.
Once we obtain the transfer function of a quantum system, 
the characteristics of the internal observables is not significant 
for the input-output relation.

Eq.(\ref{guncertain}) allows us to simplify the transfer
function of the quantum SISO linear system.
One can choose $B_x=B_p=-C_x$
with an appropriate similar transformation.
Since Eq.(\ref{guncertain}) holds for an arbitrary $s$,
it follows that $D_xD_p = 1$. 
For a physical reason, it is natural to assume $|G_x|=|G_y|$ at
infinite frequencies.
Thus, with an appropriate phase rotation, 
we have $D_x=D_p= 1$.
Furthermore, Eq.(\ref{guncertain}) leads to 
\begin{align}
 s(1-C_xC_p^{-1})-(A_x+A_p+C_x^2)=0,
\nonumber
\end{align}
which verifies that $C_x=C_p$ and $A_x+A_p=-C_x^2$.
After all, 
we can express a general linear quantum  system as
\begin{align}
& & \hspace*{-5mm}
\matAV{dm_x}{dm_y}\hspace{-1mm}=\hspace{-1mm}
\tf{\matAA{-\frac{1}{2}C_x^2+F}{0}{0}{\hspace{-5mm}-\frac{1}{2}C_x^2-F}}
   {\matAA{-C_x}{0}{0}{-C_x}}
   {\matAA{C_x \ \ }{ \ \ 0}{0 \ \ }{ \ \ C_x}}
   {\matAA{1 \ \ }{ \ \ 0}{0 \ \ }{ \ \ 1}}
\hspace{-1mm}
\matAV{du_x}{du_y},
\nonumber \\ & & \vspace{-4mm}
\label{ung}
\end{align}
where $F$ is a free parameter.
Note that this form also holds for multi-input-multi-output systems if
each matrix is nonsingular.

\section{measurements on linear systems}
\label{mp}

Let us consider a measurement of a single phase of the output, say
$m_x$, of the linear quantum system (\ref{ung}).
This is called the homodyne measurement.
The measurement projects the signals $u_x,u_y$ 
onto a commutative counterparts $\xi,\eta$
with the same stochastic properties as the quantum ones.
Since we are measuring only $x$-phase, no information on $y$-phase is
valid.
Thus, we can represent the measured system as
\begin{subequations}
\label{ql} 
\begin{align}
  d\matAV{x}{y}
  &= A \matAV{x}{y}dt 
  + B\matAV{d\xi}{d\eta},
\\
  d\underbar{m} &= C\matAV{x}{y}dt
  + D\matAV{d\xi}{d\eta},
\end{align}
\end{subequations}
where $\underbar{m}$ is the measurement outcome, 
$x,y$ are the states 
(we have used the same notations as Eq.(\ref{dft})
although they are commutative here),
and each matrix is given by 
\begin{align}
  A=\matA{-\frac{1}{2}C_x^2+F}{0}{0}{\hspace{-5mm}-\frac{1}{2}C_x^2-F}
& \hspace{1em}
  B=\matA{-C_x}{0}{0}{-C_x}
\nonumber \\ 
  C=\matA{C_x}{0}{0}{0} & \hspace{1em}
  D=\matA{1}{0}{0}{0}.
\nonumber
\end{align}
This is alternatively represented as
\begin{align}
 d\underbar{m} =
 \tf{\matAA{-\frac{1}{2}C_x^2+F}{0}{0}{\hspace{-5mm}-\frac{1}{2}C_x^2-F}}
    {\matAA{-C_x}{0}{0}{-C_x}}
    {\matAA{C_x}{0}{0}{0}}
    {\matAA{1}{0}{0}{0}}\matAV{d\xi}{d\eta}.
\nonumber
\end{align}
From the expression above, 
it is obvious that the second element of $\underbar{m}$ is irrelevant.

The projection postulate of the quantum theory means that 
{\it the quantum system under measurements obeys a density conditioned
on the measurement outcomes.}
Since for linear systems with correlated noises
the conditional density is given by Eq.(\ref{cc}), which is obtained via
the Kalman filter,
we have the conditional density of the quantum system on the phase space
\begin{align}
 dp &= \Bigl[-\frac{\p}{\p x} (-\frac{1}{2}C_x^2+F) x 
     -\frac{\p}{\p y} (-\frac{1}{2}C_x^2-F) y
\nonumber \\ &  \hspace*{3em}
 + \frac{1}{2}C_x^2\Bigl(
  \frac{\p^2}{\p x^2}
+ \frac{\p^2}{\p y^2}
\Bigr)\Bigr]pdt
\nonumber \\  & \hspace*{5em}
 + C_x\Bigl[x 
    - \mean{x} 
    - \frac{\p}{\p x}  \Bigr]
 p \ d\underbar{v},
\label{qp}
\end{align}
where $\underbar{v}$ is the innovation process.
Using the correspondences between the density matrix $\rho$ and the
density on the phase space $p$ \cite{garb}
\begin{subequations}
\label{gar} 
\begin{align}
 a\rho &\leftrightarrow 
 \Bigl(\alpha +\frac{1}{2}\frac{\p}{\p \alpha^*}\Bigr)p, 
 \\
 a\dgg\rho &\leftrightarrow 
 \Bigl(\alpha^* -\frac{1}{2}\frac{\p}{\p \alpha}\Bigr)p, 
 \\
 \rho a &\leftrightarrow
 \Bigl(\alpha -\frac{1}{2}\frac{\p}{\p \alpha^*}\Bigr)p, 
 \\
 \rho a\dgg &\leftrightarrow 
 \Bigl(\alpha^* +\frac{1}{2}\frac{\p}{\p \alpha}\Bigr)p,
\end{align}
\end{subequations}
where $a$ is the annihilation operator and $\alpha :=(x+\im y)/2$,
we obtain the quantum stochastic master equation
\begin{align}
  d\rho =& 
\frac{C_x^2}{2}(2a\rho a\dgg -a\dgg a \rho - \rho a\dgg a)dt
\label{qsm}\\ 
  &+  [F(a\dgg a\dgg - aa), \rho]dt
  +C_x(a\rho + \rho a\dgg -\mean{x}\rho)
  d\underbar{v}.
\nonumber
\end{align}
Note that the derivation of this stochastic master equation started from
Eq.(\ref{guncertain}).

The derivatives in RHS's of Eq.(\ref{gar}) come from 
the noncommutativity of the operators $a,a\dgg$, and therefore
the derivative in the innovation process
of Eq.(\ref{qp}) is
the consequence of the fact that we are measuring the
operator-valued variable. 
On the other hand, 
as the Kalman filter, the derivative results from the correlation
between the dynamical noise and the measurement noise, as stated earlier. 
Thus, the quantum effect of the measurement on linear 
quantum systems is expressed as the classical noise correlation.

Let us consider the first moments of noncommutative observables $x,y$
subject to Eq.(\ref{qsm}), 
defined as 
$\hat{x}=\Tr (a+a\dgg) \rho,  \ \hat{y}=\im \Tr (a\dgg-a)\rho$.
They are calculated as 
\begin{align}
\hspace*{-.5em}
 d\matAV{\hat{x}}{\hat{y}}
 = A\matAV{\hat{x}}{\hat{y}}dt
 +(PC^T+B) (DD^T)^{-1}\matAV{d\bar{\xi}}{d\bar{\eta}},
\label{qk}
\end{align}
where 
the covariance matrix $P$
obeys a Riccati differential equation
\begin{align}
 \dot{P} =& 
 AP+PA^T+BB^T
\label{qv} \\ & 
+(PC^T+BS)(DD^T)^{-1}(PC^T+BS)^T.
\nonumber
\end{align}
Now it is obvious that this is equivalent to Eq.(\ref{cr}).

Assume that $(C,A)$ is detactable and $(A,B)$ is stabilizable.
Then, the system is in the stationary state
after the measurement for an infinitely long time \cite{kai}, 
and Eq.(\ref{qv}) can be reduced to an algebraic Riccati equation
associated with a Hamiltonian matrix
\begin{align}
\hspace*{-.5em}
 H = \matA{\matAA{\frac{C_x^2}{2}+F}{0}{0}{-\frac{C_x^2}{2}-F}}
{\matAA{-C_x^2\hspace{5mm}}{\hspace{5mm}0}{0\hspace{5mm}}{\hspace{5mm}0}}
{\matAA{0\hspace{5mm}}{\hspace{5mm}0}{0\hspace{5mm}}{\hspace{5mm}-C_x^2}}
          {\matAA{-\frac{C_x^2}{2}-F}{0}{0}{\frac{C_x^2}{2}+F}}.
\label{ueh}
\end{align}
If $F> -C_x^2/2$,
the eigenspace corresponding to the negative eigenvalues of $H$ is
spanned by 
\begin{align}
   \left[
    \begin{array}{c} 
        1 \\           
        0 \\
        1 + \frac{2F}{C_x^2} \\
        0
    \end{array}
  \right],
\qquad
   \left[
    \begin{array}{c} 
        0 \\           
        1 \\
        0 \\
        (1 + \frac{2F}{C_x^2})^{-1}
    \end{array}
  \right].
\end{align}
According to Eq.(\ref{arim}), 
one can obtain the stationary covariance matrix $P$ given by
\begin{align}
  P = \matA{1 + \frac{2F}{C_x^2}}{0}{0}{(1 + \frac{2F}{C_x^2})^{-1}}.
\end{align}
This implies the trade-off between the variances of the two phases,
as expected from the uncertainty relation.
It should be noted that this trade-off results from Eq.(\ref{guncertain}).
If $F=0$, the system asymptotically goes to the vacuum state, which is
also physically natural.

\section{stationary uncertainty}
\label{sur}

So far, we have seen the constraints resulting from Eq.(\ref{guncertain}).
In this section, we shall consider the general case of
Eq.(\ref{guncertain1}).
For a quantum system the covariance matrix satisfies the following inequality 
\cite{hol}:
\begin{align}
 P+\Omega \ge 0,
\label{cou}
\end{align}
where 
\begin{align}
 \Omega := \matA{0}{-\im}{\im}{0}.
\nonumber
\end{align}
In the static case, $P=\mbox{Ric}[H]$ with a Hamiltonian matrix
\begin{align}
 H=\matA{(A-BSC)\dgg}{-C\dgg C}{-B(I-S^2)B\dgg}{-(A-BSC)},
\end{align}
where the transpose is generalized to the complex conjugate.

Let us consider the similarity transformation of $H$ with a matrix
\begin{align}
 T = \matA{I}{0}{\Omega}{I}.
\nonumber
\end{align}
From the definition (\ref{dha}), it turns out that $THT^{-1}$ is also a
Hamiltonian matrix.
From the definition of $\mbox{Ric}[H]$, we have
\begin{align}
 H \matAV{I}{\mbox{Ric}[H]} = \matAV{I}{\mbox{Ric}[H]} N
\nonumber
\end{align}
for a matrix $N$ such that $\mbox{Re}\lambda(N)<0$.
Premultiplication of this equation by $T$ verifies that 
\begin{align}
 THT^{-1} \matAV{I}{\mbox{Ric}[H]+\Omega} 
= \matAV{I}{\mbox{Ric}[H]+\Omega} N,
\nonumber
\end{align}
which implies that $\mbox{Ric}[THT^{-1}]=\mbox{Ric}[H]+\Omega$.

The stationary uncertainty relation (\ref{cou}) is now equivalent to the 
condition that $\mbox{Ric}[THT^{-1}]\ge 0$.
Let us define $A_T,B_T$ and $C_T$ by
\begin{align}
 THT^{-1} = \matA{A_T\dgg}{-C_T\dgg C_T}{-B_TB_T\dgg}{-A_T}.
\end{align}
From the statement on the positivity of the solution to the algebraic
Riccati equation in Sec.\ref{hm},
all linear quantum systems are subject to the condition that 
$(C_T,A_T)$ is detectable and $(A_T,B_T)$ has no uncontrollable modes
on the imaginary axis.

For example, in the case of the linear quantum system (\ref{ql}), or the
corresponding Hamiltonian matrix (\ref{ueh}), it can be easily shown
that 
\begin{align}
 THT^{-1} 
= \matA{\matAA{\frac{C_x^2}{2}+F}{-\im C_x^2}{0}{-\frac{C_x^2}{2}-F}}
{\hspace{-3mm}
 \matAA{-C_x^2\hspace{5mm}}{\hspace{5mm}0}{0\hspace{5mm}}{\hspace{5mm}0}}
{\matAA{0\hspace{5mm}}{\hspace{5mm}0}{0\hspace{5mm}}{\hspace{5mm}0}}
{\hspace{-3mm}
 \matAA{-\frac{C_x^2}{2}-F}{0}{-\im C_x^2}{\frac{C_x^2}{2}+F}}.
\nonumber
\end{align}
For $F>-C_x^2/2$, this system satisfies the condition shown above.
Moreover, in this case, $B_T=0$ reflects the fact that the system
achieves the equality of the uncertainty relation.

\section{what do we control?}
\label{wdwc}

In the classical case, the Kalman filter is used to estimate the state
of a system, and 
we usually assume that the system is not influenced by the estimator,
i.e., the evolution of the system does not depend on how we estimate the
state. 
Then, the system (\ref{ls}) 
is described by the joint density $p(x,m)$, and the Kalman
filter (\ref{cr}) is used to obtain the conditional density $p(x|m)$, which
describes our knowledge about the system.
Using the conditional expectation $\E(x|m)$,
we control the marginal density $p(x)$.

In the quantum case, 
the joint density $p(x,y)$ describes the system before measurements,
or {\it pre-measurement} system.
If we make a measurement on the system (\ref{dft}), it no longer
obeys (\ref{dft}) but the Kalman filter (\ref{qk}).
This is what the back-action of measurements means and the consequence
of the projection postulate.
In other words, the quantum system reflects our knowledge about the
system.
Thus, the innovation process can be thought of as 
the {\it back-action process} of the measurement, and our knowledge and
the quantum system itself are continuously updated by the measurement
outcome. 
There are still uncertainties on the initial state, and the quantum
system under the measurement is different from our knowledge. 
However, the stability of the Kalman filter guarantees that they
agree with each other asymptotically.
In this case, 
the target of our control using the conditional expectation $\E(x|m)$
is the Kalman filter or our knowledge itself, instead of the marginal
density $p(x)$, and consequently,
quantum control with measurements is equivalent to controlling a
nonlinear system with state feedback.

\section{examples}
\label{exps}

In this section, we will see simple examples in which the stochastic
master equation derived from a system theoretical point of view consists
with a physical derivation.

\subsection{single mode}

In the case of a cavity, the input and the output is the traveling wave
in free space and the state is a single mode inside the cavity. 
The infinitesimal evolution of the system is given by a unitary operator
\begin{align}
 U(dt) = \exp[ 
 K(a\otimes du\dgg-a\otimes \dgg du)
 ],
\nonumber
\end{align}
where 
$a$ is the annihilation operator for the cavity mode, 
$u$ is the traveling field before the interaction with the cavity,
and $K$ is a coupling constant.
The cavity system before measurements can be expressed as
\begin{align}
  \matAV{dm_x}{dm_y}=
  \tf{A}{B}{C}{D}
  \matAV{du_x}{du_y},
\nonumber
\end{align}
where $m$ is the traveling field immediately after the interaction.
Each element of the transfer function is given by
\begin{align}
 A=\matA{-\frac{K^2}{2}}{0}{0}{-\frac{K^2}{2}}, & \hspace{2mm}
 B=\matA{-K}{0}{0}{-K}, \nonumber\\
 C=\matA{K}{0}{0}{K}, & \hspace{2mm}
 D=\matA{1}{0}{0}{1}.   \nonumber
\nonumber
\end{align}
Thus, from Eq.(\ref{qsm}), the cavity under the homodyne measurement is
described by a density matrix $\rho$ satisfying 
\begin{align}
 d\rho =& \frac{K^2}{2}(2a\rho a\dgg-aa\dgg\rho -\rho aa\dgg)dt
\label{cvt} \\ & 
 +K (a\rho + \rho a\dgg - \mean{a+a\dgg}\rho)
    d\underbar{v},
\nonumber
\end{align}
where $d\underbar{v}$ is the innovation process.
This is equivalent to the standard stochastic master equation for
the optical system \cite{wis1}.

\subsection{deterministic input}

Let us consider an additional deterministic input 
to the system (\ref{cvt}) described by a Hamiltonian
\begin{align}
 H = -h_y(a+a\dgg) -\im h_x(a-a\dgg),
\nonumber
\end{align}
where $h_x,h_y$ are control gains.
The conditional expectation is then given by
\begin{subequations}
\label{linp}
\begin{align}
  d\matAV{\hat{x}}{\hat{y}} \hspace{-1mm}
  =& A\matAV{\hat{x}}{\hat{y}}dt 
  + \matAV{h_x}{h_y}dt
  + (PC^T-B)d\underbar{v}, 
 \\ 
 \dot{P} =& AP+PA + BB^T 
 \\ & 
- (PC^T+B)(DD^T)^{-1}(PC^T+B)^T.
\nonumber
\end{align}
\end{subequations}
where each matrix is given by 
\begin{align}
  A=\matA{-\frac{K^2}{2}}{0}{0}{\hspace{-5mm}-\frac{K^2}{2}}
 & \hspace{3mm}
  B=\matA{-K}{0}{0}{-K}
\nonumber \\ 
  C=\matA{K}{0}{0}{0} & \hspace{3mm}
  D=\matA{1}{0}{0}{0}.
\nonumber
\end{align}
Since the control input $h_x,h_y$ is deterministic and linear,
the covariance matrix $P$ is independent of the input.
This fact simplifies the design of the input to stabilize the Kalman
filter. 
However, this is not the case in general, as will be seen later,
and it is difficult to show the stability.

\subsection{additional noise}

Assume that a classical noise is added to the system 
during signal processing before we obtain the measurement outcome
$\underbar{m}_x$.
The whole system is described by 
\begin{align}
  d\matAV{\underbar{x}}{\underbar{y}} 
  =
  \tf{\matAA{\hspace{-1mm}-\frac{K^2}{2}}
    {0}
    {0}
    {\hspace{-2mm}-\frac{K^2}{2}}}
     {
    \begin{array}{ccc}    
        \hspace{-1mm} -K & \hspace{-2mm} 0 & 0 \\   
        0 & \hspace{-2mm} -K & 0 
    \end{array}}
     {\matAAH{K}{0}}
     {\matBBH{1}{0}{\delta}}
\matBV{d\xi}{d\eta}{d\zeta},
\label{qll}
\end{align}
where $\zeta$ is the additional Winer process independent of
$\xi,\eta$, and $\delta$ is the strength of the noise.
For this equation, $DD^T = 1+\delta^2$, so the stochastic master
equation is given by
\begin{align}
 d\rho =& \frac{K^2}{2}(2a\rho a\dgg-aa\dgg\rho -\rho aa\dgg)dt
\nonumber \\ &
 +\frac{K}{1+\delta^2} (a\rho + \rho a\dgg - \mean{a+a\dgg}\rho)
  d\underbar{v}.
\nonumber
\end{align}
This is equivalent to the stochastic master equation for the photon
detector with efficiency $1/(1+\delta^2)$, in which $\zeta$ is thought
of as a noise induced by undesirable optical modes.
In reality, we cannot discriminate between the classical and quantum
noise in the measurement process.
Eq.(\ref{cvt}) has assumed an ideal signal amplification of the photon
detector. 
If we can obtain the full dynamical model of the measurement process,
including information and noise processing in the photon detector, 
the stochastic master equation would be of a different form.
It would be described by the Kalman filter for a cascade system
(\ref{cas}).

\section{spin Wigner function}
\label{bswf}

The phase space formulation 
is compatible with the classical formulation and 
allows us to understand quantum systems on the classical language.
We have seen that 
once the transfer function representation of a quantum system 
is obtained, the stochastic master equation naturally follows from the
Kalman filter in a classical way. 
A spin system is described by a density on a sphere,
instead of the phase space, and the spherical constraint leads to the
nonlinearities of the dynamics. 
The calculation of a conditional density then leads to a general
filtering problem. 
For spins, however, there does not necessarily exist 
the classical filter corresponding to the quantum stochastic master
equation.
In this section, we will examine the classical counterparts
to spin systems under measurements.

For an operator $X$ on a Hilbert space representing an $S$-spin system, 
let us define a function as 
\begin{align}
  W_X(\Omega) = \Tr X w(\Omega),
 \label{spinw}
\end{align}
where
\begin{align}
  w(\Omega) = \sqrt{\frac{4\pi}{2S+1}}
  \sum_{L=0}^{2S}\sum_{M=-L}^{L}
  Y^*_{LM}(\Omega)T_{LM}.
\label{w}
\end{align}
$Y_{LM}(\Omega)$ are the spherical harmonics and $T_{LM}$ are 
the irreducible tensors of rank $L$, or the polarization operator,
which are, in terms of the basis spin functions, defined as
\begin{align}
  T_{LM} = \sqrt{\frac{2L+1}{2S+1}}\sum_{mm'}
  \cg{Sm}{LM}{Sm'} \ket{Sm'}\bra{Sm},
\end{align}
where $\cg{Sm}{LM}{Sm'}$ are Clebsch-Gordan coefficients.
For a density matrix $\rho$ of the spin system,
the function $W_{\rho}$ has the same properties as the
optical Wigner function. 
For example, the expectation of an arbitrary
operator $A$ with respect to $\rho$ is expressed as 
\begin{align}
 \Tr \rho A = \frac{2S+1}{4\pi} \int d\Omega \
 W_{\rho}W_{A},
\end{align}
where $d\Omega = \sin\theta d\theta d\phi$.
From $W_{\rho}$,
$\rho$ can be reconstructed via the relation
\begin{align}
 \label{ro}
  \rho = \frac{2S+1}{4\pi} \int d\Omega \
  w(\Omega)W_{\rho}(\Omega).
\end{align}

Let us define an operator $\maths{P}$ as
\begin{align}
 \maths{P}(W_AW_B) = W_{AB}.
\end{align}
For the $S$-spin system, $\maths{P}$ is given by \cite{kli}
\begin{align}
  \maths{P} =& \sum_j 
  \frac{(-1)^j}{j!(2S+j+1)!}
  \sqrt{2S+1} \tilde{F}^{-1}(\maths{L}^2)
\label{p} \\ &  \hspace*{2em}
  \times
  \Bigl\{
  \maths{R}^{+j}\tilde{F}(\maths{L}^2)
  \otimes
  \maths{R}^{-j}\tilde{F}(\maths{L}^2)
  \Bigr\},
\nonumber
\end{align}
where $\tilde{F}$ is defined by
\begin{align}
  \tilde{F}(\maths{L}^2) Y_{LM} = \sqrt{(2S+L+1)!(2S-L)!} \ Y_{LM},
\nonumber
\end{align}
and 
\begin{align}
 \maths{L}^2 &=
 -\Bigl(
 \frac{\p^2}{\p\theta^2}+\cot\theta\frac{\p}{\p\theta}
 +\frac{1}{\sin^2\theta}\frac{\p^2}{\p\phi^2}
 \Bigr),
\nonumber \\
 \maths{R}^{\pm j} &= \prod_{k=0}^{j-1}
 (k\cot\theta 
 -\frac{\p}{\p\theta}
 \mp \frac{\im}{\sin\theta}\frac{\p}{\p\phi}).
\nonumber 
\end{align}
Note that $\maths{L}^2$ and $\maths{R}^{\pm j}$ are commutative.
In the large spin limit $S\gg 1$, $\maths{P}$ can be simplified as
\begin{align}
  \maths{P} = 1\otimes 1 + \frac{\epsilon}{2}
  (\maths{R}^{-1}\otimes\maths{R}^{+1} 
- \maths{R}^{+1}\otimes\maths{R}^{-1}),
 \label{pe}
\end{align}
where $\epsilon=(2S+1)^{-1}\ll 1$.
Let us define complex numbers $\alpha,\alpha^*$ as 
\begin{subequations}
\label{aad} 
\begin{align}
 \alpha &= e^{\im\phi}\sin\theta,
 \\
 \alpha^* &= e^{-\im\phi}\sin\theta.
\end{align}
\end{subequations}
In the limit of $\theta\ll 1$, $\maths{R}^{\pm}$ can be
represented as 
\begin{align}
  \maths{R}^{-1} &\sim - e^{\im\phi} \frac{\p}{\p\alpha},
\nonumber \\
  \maths{R}^{+1} &\sim - e^{-\im\phi}\frac{\p}{\p\alpha^*}.
\nonumber
\end{align}

The spin angular momentum operator or briefly the spin operator,
can be represented by a set of three $(2S+1)\times(2S+1)$ matrices.
Using the polarization operators, 
the spherical components of the spin operator are given by
\begin{subequations}
\label{st}
\begin{align}
  S_{\pm} &= \mp \sqrt{\frac{S(S+1)(2S+1)}{3}} T_{1\pm 1},
 \\
  S_0 &= \sqrt{\frac{S(S+1)(2S+1)}{3}} T_{10},
\end{align}
\end{subequations}
and the cartesian components are
\begin{subequations}
\begin{align}
 S_x &= (S_+ + S_-),
 \\
 S_y &= -\im(S_+ - S_-),
 \\
 S_z &= S_0.
\end{align}
\end{subequations}
For the spherical components of the spin operator,
the definition (\ref{spinw}) yields
\begin{subequations}
\label{wxyz} 
\begin{align}
  W_{S_+} &= \sqrt{\frac{S(S+1)}{2}} e^{+\im\phi}\sin\theta 
           =  \sqrt{\frac{S(S+1)}{2}} \alpha,
\\
  W_{S_-} &= \sqrt{\frac{S(S+1)}{2}} e^{-\im\phi}\sin\theta 
           =  \sqrt{\frac{S(S+1)}{2}} \alpha^*,
\\
  W_{S_0} &= \sqrt{S(S+1)} \cos\theta. 
\end{align}
\end{subequations}

Let us consider the action of the spin operators on the density matrix
in terms of the spin Wigner function.
According to Eq.(\ref{pe}), we have
\begin{align}
  S_+\rho =&
  \frac{2S+1}{4\pi}\int d\Omega w \maths{P}(W_{S_+}W_{\rho})
\\
  =&
  \frac{2S+1}{4\pi}\int d\Omega w 
  \sqrt{\frac{S(S+1)}{2}}
  \Bigl(
  \alpha + \frac{\epsilon}{2}\frac{\p}{\p\alpha^*}
  \Bigr)W_{\rho}.
\nonumber
\end{align}
The other forms can be calculated in the same way, and we obtain the
following correspondences between the spin operators and 
the differential operators:
\begin{subequations}
\label{sgar} 
\begin{align}
 S_+\rho &\leftrightarrow
  \sqrt{\frac{S(S+1)}{2}}
  \Bigl(
  \alpha + \frac{\epsilon}{2}\frac{\p}{\p\alpha^*}
  \Bigr)W_{\rho},
 \\
 S_-\rho &\leftrightarrow 
  \sqrt{\frac{S(S+1)}{2}}
  \Bigl(
  \alpha^* - \frac{\epsilon}{2}\frac{\p}{\p\alpha}
  \Bigr)W_{\rho},
 \\
 \rho S_+ &\leftrightarrow 
  \sqrt{\frac{S(S+1)}{2}}
  \Bigl(
  \alpha - \frac{\epsilon}{2}\frac{\p}{\p\alpha^*}
  \Bigr)W_{\rho},
 \\
 \rho S_- &\leftrightarrow 
  \sqrt{\frac{S(S+1)}{2}}
  \Bigl(
  \alpha^* + \frac{\epsilon}{2}\frac{\p}{\p\alpha}
  \Bigr)W_{\rho}.
\end{align}
\end{subequations}
These relations are similar to the bosonic case (\ref{gar})
except for the small parameter $\epsilon$ in the second terms.

\section{local behavior of large spins}
\label{lbls}

The correspondences shown above involve the complex parameters
$\alpha,\alpha^*$.  
This fact implies that in the large spin limit, the spin operator can be
related to the creation and annihilation operators of the basonic system. 
In fact, 
the relation between the spin operator and creation and annihilation
operators can be directly seen from the commutation relation
of the spin operator.
Let us define operators $A_{\pm},A_z$ by
\begin{align}
  A_{\pm} =& \mu S_{\mp},
\nonumber \\
  A_z =& -S_z + \frac{1}{2\mu^2},
\nonumber
\end{align}
where $\mu$ is a constant.
They obey the commutation relations
\begin{align}
  [A_-,A_+] =& 1-2\lambda^2 A_z,
\nonumber \\
  [A_{\pm},A_z] =& A_{\pm}.
\nonumber
\end{align}
Thus, if $\lambda \to 0$, we have the correspondences 
\begin{subequations}
\label{soc} 
\begin{align}
  A_+ &\leftrightarrow a\dgg,
 \\
  A_- &\leftrightarrow a,
 \\
  A_z &\leftrightarrow a^+ a.
\end{align}
\end{subequations}
From Eq.(\ref{aad}), it turns out that the north pole of the sphere
corresponds to the origin of the phase space so that 
the the spin function $\ket{SS}$ is the ground state.
Thus, we have
\begin{align}
 \lim_{c\to 0} A_z \ket{SS} = 0,
\nonumber
\end{align}
which implies $S\sim (2\mu^2)^{-1}$, and consequently, 
$S\gg 1$ is required for the correspondences (\ref{soc}) to hold.

Let us consider the measurement of $x$ component of the spin operator 
with damping.
The stochastic master equation is given by
\begin{align}
 d\rho =&  \frac{K^2}{2}  
  (2S_+\rho S_- -\rho S_-S_+ - S_-S_+ \rho ) dt
\nonumber  \\  &
 + K  (S_+\rho +\rho S_- -\mean{S_++S_-}\rho) d\underbar{v},
\nonumber
\end{align}
where $\underbar{v}$ 
is the innovation process resulting from the measurement of $S_x$. 
Define two variables as
\begin{align}
  \matAV{X}{Y} = \sqrt{2S+1}
  \matA{1}{1}{-\im}{\im}\matAV{\alpha}{\alpha^*}.
\nonumber
\end{align}
Eq.(\ref{sgar})
leads to the stochastic master equation for the spin Winger function 
written as
\begin{align}
  dW_{\rho} =& 
  \frac{K^2}{2\epsilon} \Bigl[
    \frac{\p}{\p X}X + \frac{\p}{\p Y}Y
  +\frac{\p^2}{\p X^2}+\frac{\p^2}{\p Y^2}
  \Bigr]W_{\rho} dt
\nonumber
\\ &
+ \frac{K}{\sqrt{\epsilon}}\Bigl[ X + \frac{\p}{\p X} -\mean{X}
  \Bigr]W_{\rho} \ d\underbar{v}
\end{align}
This is the same as the evolution of the conditional density for the
bosonic system (\ref{qp}).
Thus, the local behavior of the large spin is equivalent to the bosonic
system, and the stochastic master equation can also be derived as 
the Kalman filter of the following pre-measurement system:
\begin{align}
 \hspace*{0em}
  d\matAV{X}{Y} 
&=
\matA{-\frac{K^2}{2\epsilon}}
    {0}
    {0}
    {\hspace{-2mm}-\frac{K^2}{2\epsilon}} 
    \matAV{X}{Y}dt 
  + \matA{-\frac{K}{\sqrt{\epsilon}}}
         {0}
         {0}
         {\hspace{-2mm}-\frac{K}{\sqrt{\epsilon}}}
  \matAV{d\xi}{d\eta},
\nonumber \\ 
  d\underbar{m}_x 
&=
  \matAH{\frac{K}{\sqrt{\epsilon}}}{0} 
  \matAV{X}{Y}dt 
  + \matAH{1}{0}  \matAV{d\xi}{d\eta}.
\nonumber
\end{align}
Each element of $A,B$ and $C$ matrices grows with the number of spins,
and consequently the measurement noise relatively becomes small accordingly.
In other words, the measurement of the large number spin is almost
deterministic.

In the case of QND measurement of 
the $x$ component of the spin, $S_x$, 
the stochastic master equation is given by \cite{tho}
\begin{align}
  d\rho = 
  \frac{K^2}{2}[S_x,[S_x,\rho]] dt 
  + K (\{S_x,\rho\}-2\mean{S_x}\rho) d\underbar{v}.
\nonumber
\end{align}
To zeroth order of $\epsilon$, 
Eq.(\ref{sgar}) yields
\begin{align}
 dW_{\rho} = 
 \frac{K^2}{4\epsilon} \frac{\p^2}{\p Y^2} W_{\rho} dt
+ \frac{K}{\sqrt{2\epsilon}}(X-\mean{X})W_{\rho} \ d\underbar{v}.
\label{xqw}
\end{align}
From Eq.(\ref{cd}), it can be easily seen that this stochastic master
equation is equivalent to 
the Kalman filter of the following pre-measurement system:
\begin{subequations}
\label{xqnd}
\begin{align}
 d\matAV{X}{Y} =& 
 \frac{K}{\sqrt{2\epsilon}} \matAV{0}{1}dw
\label{xqnd1} \\ 
 d\underbar{m}_x =& 
 \frac{K}{\sqrt{2\epsilon}}
 \matAH{1}{0}\matAV{X}{Y}dt + dv,
\label{xqnd2}
\end{align}
\end{subequations}
where $w,v$ are independent normalized Wiener processes.
These equations exactly reflect what QND measurement means.
Eq.(\ref{xqnd1}) indicates that under QND measurement of $S_x$,
the corresponding variable $X$ is not affected by the dynamical noise.
However, there is the measurement noise $v$ which comes from the
interaction between the spin and the external field for extracting
information from the system.
Because of the measurement noise,
the measured system must be described by the conditional density, 
or the Kalman filter of Eq.(\ref{xqnd}), 
and then $X$ is subject to the back-action noise
$\underbar{v}$.

\section{global behavior of spins}
\label{ghos}

In the previous section, we have seen that the local behavior of a
large spin is well-described on the local $XY$ plane at the north
pole of the sphere.
The $x$ and $y$ components of the spin is equivalent to the two phases
of a bosonic system, and thereby described as the filter of a linear
classical system. 
For the global behavior of the large spin under QND measurement,
it is convenient to use the $z$ component for a mathematical simplicity.
We assume here that,
in addition to the QND measurement,
there is a control parameter which allows us to rotate the spin 
about $y$ axis with a control gain $h$.
The stochastic master equation for the conditioned density matrix 
is given by \cite{tho}
\begin{align}
  d\rho =&
  \Bigl(
  -\im h[S_y,\rho] - \frac{K^2}{2} [S_z,[S_z,\rho]]
  \Bigr)dt
\nonumber \\
 &
  + K
  \Bigl(
  \{S_z,\rho\} - 2\mean{S_z}\rho
  \Bigr)d\underbar{v},
\label{gzq}
\end{align}
where $\underbar{v}$ is the innovation process.

Let us rewrite the stochastic master equation for the spin Wigner
function $W_{\rho}$. 
For details, see Appendix.
It can be shown that 
the commutation relation with the cartesian components of the spin
operator $S_i \ (i=x,y,z)$ corresponds to the orbital angular
momentums $\maths{L}_i$ as
\begin{align}
  [S_i,\rho] \leftrightarrow -\maths{L}_i W_{\rho}.
\nonumber
\end{align}
Unlike the commutation relation,
the anticommutation relation with $S_z$ has a complex form.
It can be given to order of $\epsilon$ by
\begin{align}
  \{S_z,\rho\} \leftrightarrow 
  \Bigl[\frac{\cos \theta}{\epsilon} 
  -\frac{\epsilon}{2}\Bigl(\cos\theta+\sin\theta\frac{\p}{\p \theta}
  +\cos\theta \maths{L}^2
  \Bigr)
  \Bigr]W_{\rho}.
\nonumber
\end{align}
The stochastic master equation is then written for the spin Wigner
function as
\begin{align}
  dW_{\rho} = &
  \Bigl[\im h\maths{L}_y-\frac{K^2}{2}\maths{L}_z^2\Bigr]W_{\rho}dt
\label{ssq}
 \\ &
  +K \Bigl[
  \Bigl(\frac{1}{\epsilon}-\frac{\epsilon}{2}\Bigr)
  (\cos\theta -\mean{\cos\theta})
\nonumber \\ &  \hspace*{3em}
-\frac{\epsilon}{2}
  \Bigl(\sin\theta \frac{\p}{\p \theta}+\cos\theta\maths{L}^2\Bigr)
  \Bigl]W_{\rho} d\underbar{v}.
\nonumber
\end{align}
It is worth noting that the innovation process includes second
order derivatives $\maths{L}^2$.
In the case of the classical nonlinear filtering,
the innovation term is involved in derivatives of at most first order,
which occurs if the dynamical and measurement noises are correlated.
Unlike the bosonic case,
the stochastic master equation for the spin system therefore has 
no classical counterpart in the first or higher order
approximation with respect to $\epsilon$, and
the derivative of second order represents a quantum effect of the
measurement process.

To order of $1/\epsilon$, 
the stochastic master equation has a classical counterpart, and it would
be insightful to see the corresponding before-measurement
system. 
One can easily show that the stochastic master equation is equivalent to
the filtering of the following pre-measurement system:
\begin{subequations}
\label{sde}
\begin{align}
  d{\bf x} = &  A{\bf x} dt
  +B {\bf x} \ dw,
\label{sde1} \\ 
  dm =& C{\bf x} dt + \sqrt{\epsilon} dv,
\label{sde2}
\end{align}
\end{subequations}
where 
\begin{align}
 A =\hspace{-1mm}
  \matB{-\frac{K^2}{2}}{\hspace{-2mm}0}{h}
              {0}{\hspace{-2mm}-\frac{K^2}{2}}{0}
              {-h}{\hspace{-2mm}0}{0}\hspace{-1mm},
  B =\hspace{-1mm}
 \matB{0}{\hspace{-2mm}-K}{0}
      {K}{\hspace{-2mm}0}{0}
      {0}{\hspace{-2mm}0}{0}\hspace{-1mm},
 C =& \matBH{0}{0}{K}\hspace{-1mm},
\nonumber
\end{align}
and ${\bf x}=[x \ y \ z]^T$ corresponds to each cartesian components
of the spin operator, and $w,v$ are independent Wiener processes.
The double commutation relation, $[S_z,[S_z,\rho]]$, of Eq.(\ref{gzq})
produces the exponential decay of $x,y$ ($A$ matrix) and mixing of $x,y$
($B$ matrix) in Eq.(\ref{sde1}).
Apart from the control input $h$, $z$ is completely static. 
The second equation shows that the measurement process would be
deterministic in the large spin limit $\epsilon\to 0$.

The measured system is described by the filtering of Eq.(\ref{sde}),
which is given by
\begin{subequations}
\label{sek}
\begin{align}
 d\hat{{\bf x}} &= A\hat{{\bf x}}dt + PC^T d\underbar{v},
\label{sek1} \\  \hspace*{-2em}
\dot{P} &= AP+PA^T +BVB^T -\frac{1}{\epsilon}PC^TCP,
\label{sek2} \\  \hspace*{-2em}
\dot{V} &= AV+VA^T +BVB^T,
\label{sek3}
\end{align}
\end{subequations}
where $P$ is the covariance matrix and $V$ is the conditional second
moment. 
Note that the control input $h$ is of the bilinear form in
Eq.(\ref{sde}), so that the covariance matrix $P$ is dependent on
the control input $h$, unlike the bosonic case of Eq.(\ref{linp}).

\section{discussion}
\label{dis}

For quantum information theoretical purposes, we want to prepare
entangled states with the use of feedback.
Entanglement can be produced by quantum mechanical ambiguities.
In the case of an ensemble of $N$ spin-1/2 particles, the coherent
superposition of states in which half of the spins are up and the rest
are down is maximally ambiguous, e.g., 
\begin{align}
 \sum_{\sigma}
 \ket{\uparrow_{\sigma(1)} \cdots \uparrow_{\sigma(\frac{N}{2})}
 \downarrow_{\sigma(\frac{N}{2}+1)} \cdots \downarrow_{\sigma(N)}},
\nonumber
\end{align}
where $\sigma$ represents permutation.
This state can be characterized by the mean values of all
components of the spin operator being zero.
In particular, $z$ component is deterministic, i.e., the variance of $z$
component of this state is also zero.
The direction of the spins in the $xy$ plane is, however, completely
indeterminant and this ambiguity is the resource of entanglement.

At first, let us examine the asymptotic behavior of the system (\ref{sek}),
especially the covariance matrix in case of $h=0$.
Assume that the initial state of the $S$-spin is a spin coherent state
pointing $x$ axis.
The initial second moment is given by
\begin{align}
 V_0 = \matB{S^2}{0}{0}
            {0}{\frac{S}{2}}{0}
            {0}{0}{\frac{S}{2}}.
\nonumber
\end{align}
Eq.(\ref{sek}) yields the stationary second moment 
\begin{align}
 V_s = \matB{\frac{S}{2}(S+\frac{1}{2})}{0}{0}
            {0}{\frac{S}{2}(S+\frac{1}{2})}{0}
            {0}{0}{\frac{S}{2}}.
\end{align}
Then, the covariance matrix $P$ is given by $\mbox{Ric}[H]$, where $H$
is defined as
\begin{align}
 H=
\left[
  \begin{array}{cccccc}
-\frac{K^2}{2} & 0&0 & 0& 0& 0 \\
0& -\frac{K^2}{2} & 0& 0&0 & 0 \\
0& 0& 0& 0& 0& -\frac{K^2}{\epsilon} \\
-\frac{S}{2}(S+\frac{1}{2})& 0& 0& \frac{K^2}{2}& 0&0  \\
0& -\frac{S}{2}(S+\frac{1}{2})&0 &0 &\frac{K^2}{2} & 0 \\
0&0 & 0&0 &0 & 0 
  \end{array}
\right].
\nonumber
\end{align}
Simple manipulation verifies
\begin{align}
 P_s = \matB{\frac{S}{2K^2}(S+\frac{1}{2})}{}{}
          {}{\frac{S}{2K^2}(S+\frac{1}{2})}{}
          {}{}{0}.
\end{align}
As expected, the variance of $z$ goes to zero and the other two elements
are equally distributed. 
Though $z$ is static in the stationary situation, it is not stable as in
Eq.(\ref{sek1}) and fluctuated by the back action during the transitional
phase.
As a result, it is not necessarily possible to obtain the maximal
entanglement by the QND measurement only.

It is expected that the maximal entanglement can be obtained by
stabilizing $z$ \cite{sto2}.
Both of the quantum system under measurements and our knowledge on the
system are described by the conditional expectation with different initial
states.
In the case of the linear system with an additive control input as in
Eq.(\ref{linp}), the covariance is independent of the input and 
the difference of the initial states converges under a certain condition
so that we can make use of linear state feedback.
This is not the case in general, and the stabilization requires 
robustness for the difference of the initial state, nonlinear feedback
with higher order correlations and so on.

For the system (\ref{sek}), stabilization would be simplified using
switching control.
In addition, the design freedom is increased by 
introducing dynamics into the controller such that $dh = Qdt$, where $Q$
is a function.
The stability is examined due to a stochastic analog of Lyapunov theory 
\cite{kus}. 
We focus on the stabilization of $z$ and $h$ so that it would be enough
to consider a Lyapunov function of the form $U=U(z,h)$. 
The infinitesimal operator is given by
\begin{align}
 \mathfrak{L} 
 = -hx\frac{\p }{\p z} + Q\frac{\p }{\p u} 
   +\frac{P_{zz}^2}{2} \frac{\p^2 }{\p u^2}.
\nonumber
\end{align}
For $U=z^2+h^2$, $\mathfrak{L}U=h(-2xz+Q)+P_{zz}^2$. 
Since $\epsilon\ll 1$, Eq.(\ref{sek2}) leads to 
$\dot{P_{zz}}\sim -P_{zz}^2/\epsilon$.
Hence, a controller $Q=2xz-kh$ with $k$ such that
$k>(P_{zz}/h)^2$ can drive $z$ close to zero.
Then, the local description introduced in Sec.\ref{lbls} is valid and 
we can utilize the linear control introduced in Sec.\ref{exps}.

\section{Conclusion}

In this paper, we started the formulation of quantum systems from
the general treatment of linear systems with the noncommutative input
and output. 
The transfer function representation provides the model of the quantum
system not dependent on the commutativity of the internal observables.
We obtain the quantum mechanical constraints on the model from the
commutation relation of the input and the output.
This constraint can be characterized in two different ways.
One is the fact that the zeros and poles of the transfer functions for
the two phase are distributed symmetrically in the complex plane.
This characterization provides important information for quantum system
synthesis because the properties of the system are completely determined
by the poles and zeros.
The other is concerned with the positivity of the solution to the
algebraic Riccati equation. 
According to this constraint, there does not exist a quantum system
which does not satisfies a certain detectability condition. 

We have shown that the quantum stochastic master equation naturally
follows from the conditional density, or the Kalman filter, of the system
satisfying the conditions shown above. 
Then, the innovation process of the Kalman filter can be thought of as
the back action of the measurement.
These are also true for the large spin system to zeroth order of 
the number of spins. 
If we incorporate with higher order approximations, the spin system
reveals the quantum effect to which there are no classical counterparts.

\section*{Appendix : Proof of Eq.(\ref{ssq})}

Let us consider the commutation relation $[S_y,w]$ first.
From the definitions (\ref{w},\ref{st}), we have
\begin{align}
 [S_y,w] = -\im\sqrt{\frac{4\pi}{2S+1}}\sum_{LM}Y_{LM}^*
           [S_+-S_-,w],
\nonumber
\end{align}
where 
\begin{align}
& \hspace*{-1em}
 [S_+ - S_-,w] =
\nonumber \\ &
 \sqrt{\frac{L(L+1)}{2}}
 \Bigl(
 \cg{LM}{11}{LM+1}T_{LM+1}-\cg{LM}{1-1}{LM-1}T_{LM-1}
 \Bigr).
\nonumber
\end{align}
Each Clebsch-Gordan coefficient in this equation is calculated as
\begin{align}
  \cg{LM}{11}{LM+1} &= 
  -\sqrt{\frac{L^2-M^2+L-M}{2L(L+1)}},
\nonumber \\
  \cg{LM}{1-1}{LM+1} &= 
  \sqrt{\frac{L^2-M^2+L+M}{2L(L+1)}}.
\nonumber
\end{align}
Thus, the commutation relation turns out to be
\begin{align}
&\hspace*{-7em}
 [S_y,w] = 
\nonumber \\
\frac{\im}{2} \sqrt{\frac{4\pi}{2S+1}}\sum_{LM} T_{LM}
\Bigl(&
\sqrt{L(L+1)-M(M-1)}Y_{LM-1}^*
\nonumber \\ 
-&
\sqrt{L(L+1)-M(M+1)}Y_{LM+1}^*
\Bigr)
\nonumber
\end{align}
Using the differential relation of spherical harmonics
\begin{align}
 \frac{\p Y_{LM}^*}{\p\theta}
 \pm& M\cot\theta Y_{LM}^* = 
\nonumber \\
& \pm\sqrt{L(L+1)-M(M\mp 1)}Y_{LM\mp 1}^*e^{\mp\im\phi},
\nonumber
\end{align}
the commutation relation can be rewritten as
\begin{align}
[S_y,w] =&
 \frac{\im}{2}\sqrt{\frac{4\pi}{2S+1}}\sum_{LM} T_{LM}
 \Bigl\{
 e^{\im\phi}
 \Bigl(\frac{\p}{\p\theta}+\im\cot\theta\frac{\p}{\p\theta}\Bigr)
\nonumber \\
& \hspace*{7em}
 +e^{-\im\phi}
 \Bigl(\frac{\p}{\p\theta}-\im\cot\theta\frac{\p}{\p\theta}\Bigr)
\Bigr\} Y_{LM}^*,
\nonumber\\
= & -L_yw,
\label{a1}
\end{align}
where $L_y$ is an orbital angular momentum operator in the cartesian
components.
That is to say, for the operator $w$, the commutator with respect to the
spin operator $S_y$ can be replaced by the multiplification 
of the orbital angular momentum operators. 
For $[S_z,w]$, 
the commutation relation
\begin{align}
  [S_z,T_{LM}]=MT_{LM},
\nonumber
\end{align}
leads to 
\begin{align}
  [S_z,w] &= \sqrt{\frac{4\pi}{2S+1}}\sum_{LM}MY^*_{LM}T_{LM}
\nonumber \\
  &= -\im \frac{\p}{\p\phi}w
\nonumber \\
  &= -\maths{L}_z w.
\label{a2}
\end{align}

For the anticommutation relation $\{S_z,w\}$, 
let us start from the definitions
(\ref{w},\ref{st}) again, i.e.,
\begin{align}
 \{S_z,w\} = \sqrt{\frac{4\pi S(S+1)}{3}}
 \sum_{LM}Y^*_{LM}\{T_{10},T_{LM}\}.
\nonumber
\end{align}
The anticommutator of the irreducible tensors is given by
\begin{align}
  \{T_{10},T_{LM}\} =& \sqrt{3(2+1)}\sum_{L'}
  ((-1)^{L'}-(-1)^L)
\nonumber \\ & 
  \wig{L}{1}{L'}{S}{S}{S}
  \cg{LM}{10}{L'M}T_{L'M},
\nonumber
\end{align}
in which Clebsch-Gordan coefficients are given by
\begin{align}
  \cg{LM}{10}{L+1 M} =& \sqrt{\frac{(L+M+1)(L-M+1)}{(2L+1)(L+1)}},
\nonumber \\
  \cg{LM}{10}{L+1 M} =&-\sqrt{\frac{(L+M)(L-M)}{(2L+1)L}},
\nonumber
\end{align}
and the $6j$-symbols are 
\begin{align}
 & 
  \wig{L}{1}{L+1}{S}{S}{S} 
\nonumber \\ & 
  =
  \frac{(-1)^{L+1}}{2}
  \sqrt{\frac{(2S+L+2)(2S-L)(L+1)}{(2L+3)(2L+1)S(S+1)(2S+1)}},
\nonumber \\
 &
 \wig{L}{1}{L-1}{S}{S}{S} 
\nonumber \\ &  
  =
  \frac{(-1)^{L}}{2}
  \sqrt{\frac{(2S+L+2)(2S-L+1)L}{(2L+1)(2L-1)S(S+1)(2S+1)}}.
\nonumber
\end{align}
In addition, 
Using the differential relation of spherical harmonics
\begin{align}
  \sin\theta &\frac{\p}{\p\theta}Y_{LM} 
\nonumber \\ &  
  =
  L\cos\theta Y_{LM} 
  - \sqrt{\frac{2L+1}{2L-1}(L^2-M^2)}Y_{L-1 M} 
\nonumber \\ & 
  = 
  -(L+1)\cos\theta Y_{LM} 
\nonumber \\ &  \hspace*{4em}
  + \sqrt{\frac{2L+1}{2L+3}((L+1)^2-M^2)}Y_{L+1 M},
\nonumber
\end{align}
one can rewrite the anticommutator as
\begin{align}
& \hspace{-1em}
  \{S_z,w\}
  =
  \sqrt{\frac{4\pi}{2S+1}}\sum_{LM}T_{LM}
\nonumber \\  & 
  \frac{1}{2L+1}\Bigl[
  \sqrt{(2S+1)^2-L^2}
  \Bigl(
  L\cos\theta-\sin\theta\frac{\p}{\p\theta}
  \Bigr)
\nonumber \\  & \hspace{-1em}
  +\sqrt{(2S+1)^2-(L+1)^2}
  (L+1)\cos\theta+\sin\theta\frac{\p}{\p\theta}
  \Bigr]Y^*_{LM}
\nonumber 
\end{align}
Expanding this equation to order of $\epsilon$, we have
\begin{align} 
  \{S_z,w\}
  =&
  \sqrt{\frac{4\pi}{2S+1}}\sum_{LM}T_{LM}
\label{a3} \\ & \hspace*{-3em}
  \Bigl[
  \frac{\cos\theta}{\epsilon}
  -\frac{\epsilon}{2}\cos\theta
  (L(L+1)+1)
  -\frac{\epsilon}{2}\sin\theta\frac{\p}{\p\theta}
  \Bigr]Y^*_{LM}
\nonumber \\ 
  =&
  \Bigl[
  \frac{\cos\theta}{\epsilon}
  -\frac{\epsilon}{2}\cos\theta(\maths{L}^2+1)
  -\frac{\epsilon}{2}\sin\theta\frac{\p}{\p\theta}
  \Bigr]w,
\nonumber
\end{align}
where we have used the property of the total angular momentum 
$\maths{L}^2Y_{LM} = L(L+1)Y_{LM}$.
Eqs.(\ref{a1},\ref{a2},\ref{a3}) in combination with 
the definition (\ref{ro}) verify Eq.(\ref{ssq}).


\bibliography{apssamp}

\end{document}